\documentstyle{article}

\begin{document}
\title{\Huge Real and $p$-Adic Aspects of Quantization of Tachyons}

\author{\bf Goran S. Djordjevi\'c and Ljubi\v sa Ne\v si\'c \\
\it Department of Physics, University of Nis, Serbia and Montenegro\\
E-mail: nesiclj, gorandj@junis.ni.ac.yu}

\date{}

\maketitle

\noindent{\bf Abstract. }{A simplified model of tachyon matter in
classical and quantum mechanics is constructed. $p$-Adic path
integral quantization of the model is considered. Recent results
in using $p$-adic analysis, as well as perspectives of an adelic
generalization, in the investigation of tachyons are briefly
discussed. In particular, the perturbative approach in path
integral quantization is proposed.}

\section{Introduction}

An increasing number of researchers are testing again an (almost
twenty year) old idea that $p$-adics and $p$-adic string theory
\cite{vol} can be useful in attempts to understand ordinary
string theory, D-brane solutions and various aspects of tachyons
{\cite{sen,ghos-kaw}. Possible cosmological implications
\cite{DjDNV} are also an interesting matter.

$p$-Adic strings have many properties similar to ordinary strings,
but $p$-adic ones are much simpler. For example, one can find an
exact action for this theory, as well as for practically all
variations of $p$-adic string theory. In addition, it turns out
that $p$-adic string could be a very useful model for testing
Sen`s conjecture on tachyon condensation in open string field
theory (see e.g. Ref. 5). Corresponding $p$-adic classical
solution in string field theory can be explicitly found.

An important fact should be also considered, that is, $p$-adic
string field theory in the $p\rightarrow1$ limit reduces to the
tachyon effective action \cite{geras-shat}. It might be the case
that results for $p$-adic strings are applicable to boundary
string field theory. The next reason for further investigation is
similarity between $p$-adic D-branes (without strong coupling
problems) and the results found in vacuum cubic string filed
theory. It is interesting that the effective energy-momentum
tensor is equivalent to that of nonrotating dust \cite{sen1}.

Generally speaking, after almost twenty years of ``discovering``
$p$-adic strings, our understanding of physics on $p$-adic spaces
is still very poor. From many point of view, including pedagogical
one, it is very useful to understand (quantum) mechanical
analogies of new, unfamiliar objects, including tachyons.
Considering ``$p$-adic tachyons`` we stress results in foundation
of $p$-adic quantum mechanics, $p$-adic quantum cosmology and
their connection with standard theory on real numbers (and
corresponding spaces: Minkowsky, Riemman ...) and adelic quantum
theory.

It has been noted that path integrals are extremely useful in this
approach. Following S. Kar's \cite{kar} idea on the possibility of
the examination of zero dimensional theory of the field theory of
(real) tachyon matter, we extend this idea to the $p$-adic case.
In addition, we note possibilities for adelic quantum treating of
tachyon matter.

After some mathematical background in Section 2, we present
$p$-adic path integrals in Section 3. Section 4 is devoted to
$p$-adic strings and tachyons. Simple quantum mechanical analog of
$p$-adic tachyons is considered in Section 5. The paper is ended
by a short conclusion and suggestion for further research.

\section{$p$-Adic Numbers and Related Analysis}

Let us recall  that all numerical experimental results belong to
the field of rational numbers $Q$. The completion of this field
with respect to the standard norm $|\ |_\infty$ (absolute value)
leads to the field of real numbers $R\equiv Q_\infty$. Completion
of $Q$ with respect to the $p$-adic norms yields the fields of
$p$-adic numbers $Q_p$ ($p$ is a prime number). Each non-trivial
norm (valuation) on $Q$, due  to the Ostrowski theorem, is
equivalent either to a $p$-adic norm $|\ |_p$ or to the absolute
value function.

Any $p$-adic number $x\in Q_p$ can be presented as an expansion
\cite{vvz}
\begin{equation}
x= x^\nu(x_0+x_1p+x_2p^2+\cdots ),\quad\nu\in Z, \label{1.1}
\end{equation}
where $x_i=0,1,...,p-1$. $p$-Adic norm of  any term $x_ip^{\nu+i}$
in (\ref{1.1}) is $p^{-(\nu+i)}$. The $p$-adic norm is the
nonarchimedean (ultrametric) one. There are a lot of exotic
features of $p$-adic spaces. For example, any point of a disc
$B_\nu (a)=\{x\in Q_p:|x-a|_p\leq p^{\nu}\}$ can be treated as its
center. It also leads  to the total disconnectedness of $p$-adic
spaces.

For the foundation of the path integral approach on $p$-adic
spaces it is important to stress that no natural ordering on $Q_p$
exists. However, one can define a linear order as follows: $x<y$
if $|x|_p<|y|_p$, or when $|x|_p=|y|_p$, when there is an index
$m\geq 0$ such that the following is satisfied: $x_0=y_0,
x_1=y_1,...,x_{m-1}=y_{m-1}, x_m<y_m$ \cite{zele}. Generally
speaking, there are two analysis over $Q_p$. One of them is
connected with map $\phi: Q_p\rightarrow Q_p$, and the second one
is related to the map $\psi : Q_p\rightarrow C$.

In the case of $p$-adic valued function, derivatives of $\phi (x)$
are defined as in the real case, using $p$-adic norm instead of
the absolute value. $p$-Adic valued definite integrals are defined
for analytic functions
\begin{equation}
\label{1.2} \phi(t)=\sum_{n=0}^\infty \phi_n t^n, \qquad
\phi_n,t\in Q_p,
\end{equation}
 as follows:
\begin{equation}
\label{1.3} \int_a^b \phi(t)
dt=\sum_{n=0}^{\infty}\frac{\phi_n}{n+1} (b^{n+1}-a^{n+1}).
\end{equation}

In the case of mapping $Q_p\rightarrow C$, standard derivatives
are not possible, and several different types of
pseudodifferential operators have been introduced
\cite{vvz,ddjn}. Contrary, there is a well defined integral with
the Haar measure. Of special importance is Gauss integral

\begin{equation}
\label{1.9} \int_{Q_p}\chi_p(a x^2+b x)dx=
\lambda_p(a)|2a|_p^{-1/2} \chi_p \left(-\frac{{b^2}}
{{4a}}\right), \enskip a\neq 0,
\end{equation}
where \noindent $\chi_p(u) = \exp(2\pi i\{u\}_p)$ is a $p$-adic
additive character, and $\{u\}_p$ denotes the fractional part of
$u\in Q_p$. $\lambda_p(\alpha)$ is an arithmetic, complex-valued
function {\cite{vvz}}. To explore the existence of a vacuum state
in $p$-adic quantum theory we need
\begin{equation}
\label{2.17} \int_{Z_p}\chi_p(a x^2+b x)dx =
\left\{\begin{array}{ll}
\Omega(|b|_p), & |a|_p\leq1, \\
\frac{\lambda_p(a)}
{|2a|^{1/2}_p}\chi_p\left(-\frac{b^2}{4a}\right)
\Omega\left(\left|\frac{b}{2a}\right|_p\right),
 & |4a|_p>1,
\end{array}
\right.
\end{equation}
where $Z_p$ is the ring of $p$-adic integers ($Z_p=\{x\in Q_p:
|x|_p\leq1\}$) and $\Omega$ is the characteristic function of
$Z_p$. It should be noted that $\Omega$ is the simplest vacuum
state in $p$-adic quantum theory.

There is quite enough similarity between real numbers  and
$p$-adics and the corresponding analysis for the so called adelic
approach in mathematics \cite{ggp} and physics (see e.g. Ref. 4),
that in a sense unifies real and all $p$-adic number fields.

\section{ Path Integral in Ordinary and $p$-Adic Quantum Mechanics}

According to Feynman's idea \cite{fey}, quantum transition from a
space-time point $(x',t')$ to a space-time point $(x'',t'')$ is a
superposition of motions along all possible paths connecting
these two points. Let us remind the corresponding probability
amplitude is
$\langle x^{\prime\prime},t^{\prime\prime}|x^\prime,t^\prime
\rangle = \sum_q e^{\frac{2\pi i}{h}S[q]}$.
Dynamical evolution of any quantum-mechanical system, described by
a wave function $\psi(x,t)$, is given by
\begin{equation}
\psi(x'',t'')= \int_{Q_\infty} K(x'',t'';x',t')\psi(x',t')dx',
\label{3.2}
\end{equation}
where $  K(x'',t'';x',t')$ is a kernel of the unitary evolution
operator $U(t'',t')$. In Feynman's formulation of quantum
mechanics, $K(x'',t'';x',t')$ was postulated to be the path
integral
\begin{equation}
\label{3.3}
  K(x'',t'';x',t')= \int_{(x',t')}^{(x'',t'')} \mbox{exp}
\left( \frac{2\pi i}{h} \int_{t'}^{t''}L(q,\dot q,t)dt \right) Dq,
\end{equation}
where $x''=q(t'')$ and $x'=q(t')$.

As we know, for a classical action $\bar S(x'',t'';x',t')$, which
is a polynomial quadratic in $x''$ and $x'$, the corresponding
kernel $K$ (for one-dimensional quantum system) reads
\begin{equation}
  K(x'',t'';x',t')= \left( \frac{i}{h} \frac{\partial^2\bar
S}{\partial x''\partial x'} \right)^{\frac{1}{2}} \mbox{exp}
\left( \frac{2\pi i} {h}\bar S(x'',t'';x',t') \right). \label{3.7}
\end{equation}
It can be rewritten in the form more suitable for generalization
(at least from the number theory point of view)
\begin{equation}
  K_\infty(x'',t'';x',t') = \lambda_\infty \left(
-\frac{1}{2h} \frac{\partial^2\bar S}{\partial x''\partial x'}
\right) \left| \frac{1}{h} \frac{\partial^2\bar S}{\partial
x''\partial x'} \right|_\infty^{1/2} \chi_\infty \left( - \frac{1}
{h}{\bar S} \right), \label{3.8}
\end{equation}
where $\sqrt {ia}=\sqrt {i\mbox{sign}a\
|a|_\infty}=|a|_\infty^{1/2}\lambda_\infty(-a)$. In  (\ref{3.8}),
$\chi_\infty (a) = \exp (-2\pi i a)$ is an additive character of
the field of real numbers $R$. $D$-dimensional generalization of
the transition amplitude is:
$$   K_\infty(x'',t'';x', t) = \lambda_\infty\left( \det
\left(- \frac{1}{2h} \frac{\partial^2{\bar S}}{\partial
x_a^{''}\partial x_b'} \right)\right)     \left|  \det \left(-
\frac{1}{h} \frac{\partial^2{\bar S}}{\partial x_a^{''}\partial
x_b'} \right) \right| ^{1/2}_\infty $$
\begin{equation}
\times\chi_\infty \left( -\frac{1}{h} \bar S(x'',t'';x',t')
\right), \label{3.9}
\end{equation}
where $\lambda_{\infty}$ is defined as
\begin{equation}
\lambda_\infty\left( \det \left(- \frac{1}{2h}
\frac{\partial^2{\bar S}}{\partial x^{''}_a\partial x'_b}
\right)\right) = \sqrt{ \frac{1}{i^D}  sign  \det \left(-
\frac{1}{2h} \frac{\partial^2{\bar S}}{\partial x^{''}_a\partial
x'_b} \right)}, \label{3.10}
\end{equation}
and $x=(x_a), \quad a=1,2,\cdots, D$. By defining $\lambda_\infty
(0) =1$, one can see that this $\lambda_\infty$-function satisfies
the same properties as $\lambda_p$.

In  $p$-adic quantum mechanics dynamical differential equation of
the Schr\"odinger type does not exist and $p$-adic quantum
dynamics is defined by  the kernel $K_p(x'',t'';x',t')$ of the
evolution operator:
\begin{equation}
\psi_p(x'',t'')=U_p(t^{\prime\prime},t^{\prime})\psi_p(x^\prime,t^\prime)=
\int_{Q_p} K_p(x'',t'';x',t')\psi_p(x',t')dx'. \label{4.1}
\end{equation}
All general properties which hold for the kernel
$K(x'',t'';x',t')$ in standard quantum mechanics also hold in
$p$-adic case, where integration is now over $Q_p$. $p$-Adic
generalization of (\ref{3.3}) for a harmonic oscillator was done
in \cite{zele} starting from
\begin{equation}
K_p(x'',t'';x',t')= \int_{(x',t')}^{(x'',t'')} \chi_p \left(
-\frac{1}{h} \int_{t'}^{t''}L(q,\dot q,t)dt \right) \prod_tdq(t)
\label{4.2}
\end{equation}
($h\in Q$ and $q,t\in Q_p$).  In (\ref{4.2}) $dq(t)$ is the Haar
measure and $p$-adic path integral is the limit of a multiple Haar
integral. This approach was extended in Ref. 14.

A rather general path integral approach, valid for analytical
classical solutions, was developed for quadratic $p$-adic quantum
systems in Ref. 15 (in one-dimension)
\begin{equation}
\label{3.14}  K_p(x'',t'';x',t')= \lambda_p \left( - \frac{1}{2h}
\frac{\partial^2{\bar S}}{\partial x''\partial x'} \right) \left|
\frac{1}{h}\frac{\partial^2{\bar S}}{\partial x''\partial x'}
\right|_p^{\frac{1}{2}} \chi_p(-\frac{1}{h} {\bar S}
(x'',t'';x',t'))
\end{equation}
 and Ref. 16
(two-dimensional case).

The obtained $p$-adic result (\ref{3.14}) has the same form as
(\ref{3.10}) in the real case. The higher-dimensional $p$-adic
kernel was also considered \cite{banja}. Considering
real-ordinary and all $p$-adic  quantum mechanics on the same
foot, adelic formulation is also possible \cite{idaqp}. It could
be a good starting point to consider quantum mechanical analog of
``real``, ``$p$-adic`` and, possibly, ``adelic`` tachyons.

\section{$p$-Adic Strings and Tachyons}

The $p$-adic {\bf open} \rm string theory can be deduced from
ordinary bosonic open string theory on a D-brane by replacing the
integral over the real worldsheet by $p$-adic integral \cite{fw}.

A tachyon was defined as a particle that travels faster than
light, and consequently has negative $mass^2$. Surely, it is not a
convincing case for the tachyon. Quantum field theory offers a
much better framework for considering such a pretty exotic
physical model. If we would carry out perturbative quantization of
the scalar field by expanding the potential around $\phi=0$, and
ignore higher (cubic, ...) terms in the action, we would find a
particle-like state with $mass^2=V''(0)$. In the case of
$V''(0)<0$ we have again a particle with negative $mass^2$, {\it
i.e.}, a tachyon. The physical interpretation is that the
potential $V(\phi)$ has a maximum at the origin and hence a small
displacement of $\phi$ will make it grows exponentially. It is
associated with the instability of the system and a breakdown of
the theory.

Conventional formulation of string theory uses a first quantized
formalism. In this formulation one can get a state-particle with
negative $mass^2$, i.e. tachyons. The simplest case appears in 26
dimensional bosonic string theory. This approach is,
unfortunately, not suitable for testing tachyon`s solutions
\cite{sen}, but there are superstring theories defined in (9+1)
dimensions that have tachyon free closed string spectrum. In
addition, some string theories contain open string excitations
with appropriate boundary conditions at the two ends of the
string. So, one can ask: is there a stable minimum of the tachyon
potential around which it is possible to quantize the theory. In
the last few years there many papers devoted to this problem have
made some progress, but we will not consider them in this paper.

In $p$-adic string theory all tree level amplitudes involving
tachyons in the external states can be computed. The $p$-adic
(open) string theory is obtained from ordinary bosonic (open)
string theory on a D-brane by replacing the integral \cite{fw}
over the real world-sheet coordinates by $p$-adic integral
associated with a prime number $p$. There have been somewhat
different approaches, but we will not consider the constructions
of all these theories.

Let us see the exact effective action for the $p$-tachyon field.
It is described by the lagrangian \cite{bfow}

\begin{equation}
\label{p-lang}
 L_p=-\frac{1}{2} \phi p^{{-1}/{2}\Box}\phi +
\frac{1}{{p+1}} \phi^{p+1}.
\end{equation}

This form, obtained by computing Koba-Nielsen amplitudes for a
prime $p$, makes sense for all ({\it integer}) values of $p$. The
classical  equation of motion derived from (\ref{p-lang}) is

\begin{equation}
p^{-1/2\Box} \phi = \phi^p.
\end{equation}

Besides the trivial constant solutions $\phi=0,1$, a soliton
solution is admitted. The equations separate in the arguments and
for any spatial direction $x$ we get
\begin{equation}
\phi(x)=p^{1/2(p-1)} \exp\left(-\frac{p-1}{2p \ln p} x^2\right),
\end{equation}
a gaussian lump whose amplitude and spread are correlated
\cite{ghos-sen}.


\section{Quantum Mechanical Analogue of Tachyon Matter}

Now, we will concentrate on a relatively new field theory - the
field theory of tachyon matter was proposed by Sen  a few years
ago \cite{sen2}. The derivation of its action is based on a
rather involved argument. The obtained form is pretty strange and
different from the actions we used to be familiar ones

\begin{equation}
\label{action2}
 S = -\int d^{n+1}x V(T)\sqrt{1+\eta^{ij}\partial_i
T\partial_j T}.
\end{equation}

Let us consider $p$-adic analogue of the above action, originally
considered as real one, {\it i.e.} $\eta_{00}=-1,\
\eta_{\mu\nu}=\delta_{\mu\nu}$, where $\mu,\nu =1,2,...,n$. $T(x)$
is a $p$-adic scalar tachyon field and $V(T)$ is the tachyon
potential: $V(T)=\exp({{-\alpha T}/{2}})$. In the bosonic case,
$n=25$, $\alpha=1$, and for superstring $n=9$, $\alpha =\sqrt{2}$.
The square root appearing in action (\ref{action2}) (and its
multiplication to tachyon potential) makes this theory so unusual.

Here we examine a lower (zero-dimensional) mechanical analogue of
the field theory of $p$-adic tachyon matter (whatever it would
physically mean). As usually, the correspondence can be obtained
by the correspondence $x^i\rightarrow t, \ T\rightarrow x,\
V(T)\rightarrow V(x)$. The corresponding zero-dimensional action
reads

\begin{equation}
S_0 = -\int dt V(x) \sqrt{1-{\dot x}^2},
\end{equation}
where integration has to be performed over $p$-adic time. From the
above action it is not difficult to get the classical equation of
motion

\begin{equation}
\ddot x + f(x){\dot x}^2 = f(x),
\end{equation}
where function $f(x)$ denotes
$$f(x) = -\frac{1}{V}\frac{\partial V}{\partial x}.$$ Partial
differentiation of $p$-adic valued function is well defined,
although in this case it can be replaced by the ordinary one,
because $V=V(x)$.

Keeping in mind that exponential $p$-adic function (the tachyon
potential $V(x)=\exp({-\alpha x}$) should be understood as an
analytic function with corresponding radius of convergence
\cite{vvz} $r\sim {1}/{p}$, we obtain as in the real case
\begin{equation}
\ddot x +\alpha {\dot x}^2 = \alpha.
\end{equation}
By the replacements  $\dot x = \gamma \dot y,\ \alpha
\gamma=\frac{\beta}{m}$ and $\frac{\alpha}{\gamma}=g$ we get the
equation of motion
\begin{equation}
\label{eqmot}
 m\ddot y + \beta {\dot y}^2 = mg,
\end{equation}
which describes motion of a particle of mass $m$ moving in a
constant (say gravitational, Newtonian) field with quadratic
friction. It is interesting that this equation can be derived from
the ($p$-adic) action

\begin{equation}
\label{action1}
 S_0^{(p)} = -\int dt e^{-\frac{\beta y}{m}}
\sqrt{1-\frac{\beta}{mg}{\dot y}^2}.
\end{equation}

Surprisingly or not, the zero dimensional analog of the (Sen's)
field theory of tachyon matter offers an action integral
formulation for the system under gravity in the presence of
(quadratic) damping. The solution of the equation of motion
(\ref{eqmot}) reads

\begin{equation}
y=y_0+\frac{m}{2\beta} \ln \left(\frac{g-\frac{\beta}{m}v_0^2}
{g-\frac{\beta}{m}v^2}\right),
\end{equation}
with initial $t=0$ conditions for position $y(0)=y_0$ and velocity
$v(0)=v_0$. This solution has the same form in the real and
$p$-adic case, but the radius of convergence is rather different.

Faced with the increasing interest in various aspects of tachyon
field theory, including its $p$-adic aspect, this connection with
the field theory of tachyons through action integral formulation
seems worth mentioning and examining in general. Also,
quantization of the theory in path integral language might be very
useful and, as we know, very general (for real, $p$-adic and
adelic path integrals see, e.g. Ref. 17). However, a kernel of the
operator of evolution that corresponds to the action
(\ref{action1}) is still unknown, even in the real case. Because
of that the square root and exponential for small $\beta$ should
be expanded. If we treat $\beta$ $p$-adicaly small, in respect to
$p$-adic norm, we obtain

\begin{equation}
S_0 \sim -\int dt \left (1-\frac{\beta y}{m}
-\frac{\beta}{2mg}{\dot y}^2\right ).
\end{equation}

We have already calculated the path integral for the particle in
constant external field \cite{pri}. Here we have a slightly
changed form ($y^{\prime\prime}=y(\tau),\ y^{\prime}=y(0),\ h=1$)
$$K_p(y^{\prime\prime},\tau;y^\prime,0)=
\lambda_p\left(\frac{\beta}{mg\tau}\right)\left|\frac{mg\tau}{2\beta}
\right|_p^ {-{1\over2}} $$
\begin{equation} \times
\label{kernel} \chi_p
\left[-{\beta(y^{\prime\prime}-y^\prime)^2\over
{mg\tau}}-(\frac{\beta(y^{\prime\prime}+y^\prime)}{2m}-1)\tau+\frac{1}{48}\frac{\beta}{m}g\tau^3\right].
\end{equation}

It makes it possible to check the existence of the simplest
tachyonic vacuum state (invariant in respect to the evolution
operator), of the corresponding quantum mechanical model, {\it
i.e.}
\begin{equation}
\Omega(|y^{\prime\prime}|_p)=\int_{|y^{\prime}\leq1}
K_p(y^{\prime\prime},\tau;y^\prime,0)\Omega(|y^\prime|_p)dy^\prime.
\end{equation}
Using (\ref{2.17}) and (\ref{kernel}) we find that for the
existence of the ``ground`` $(\Omega)$ state of (quantum) $p$-adic
tachyons (here some technical details and case $p$=2 are omitted)
the following is necessary $\left|{\beta\over
m}\left(2y^{\prime\prime}\over
{g\tau}-{\tau\over2}\right)\right|_p\leq1$ for
$\left|{\beta\over{mg\tau}}\right|\leq1$, {\it or}
$\left|2y^{\prime\prime} - {g\tau^2\over2}\right|_p\leq1$ for
$\left|{\beta\over{mg\tau}}\right|_p>1$. Possible physical
implication on constraints for quantities related to the starting
tachyon action (\ref{action2}) will be discussed elsewhere. The
existence of $\Omega$ state opens the ``door`` for further adelic
generalization and investigation of higher-excited states.

As in the real case \cite{kar} a quadratic damping effect could
enter explicitly into the play treating it as a perturbation over
classical solution of the equation for a particle in constant
external field without friction. Damping effect would be
ascertained and understood through its dependence of $\beta$ term
\cite{godo} .

\section{Conclusion}

In this paper we show that quantum mechanical simplification of
the tachyon field theory, besides the real case, is possible in a
$p$-adic context. Also, an adelic generalization looks possible,
{\it i.e.} without some obvious principal obstacles. Path integral
formulation of zero-dimensional $p$-adic tachyons has been done
and some ''minimal'' conditions for their existence have been
found. Of course, how much this approach could be useful for
deeper understanding of the whole string theory and of its tachyon
sector requires time and further, in-depth research. The fact,
that the exact effective tachyon action in the usual string theory
is not known, while in $p$-adic string theory it is, is quite
enough motivation for this and similar investigations.

We would propose a few promising lines for further investigation.
The exact formula for quadratic quantum $p$-adic systems in two
and more dimensions \cite{banja} could be useful for
multidimensional generalization of $p$-adic tachyons. It is
tempting to extend our approach to $1+1$ dimensional field
theories, even nonlinear field theories would be here quite
nontrivial problem.

Finally, $p$-adic string theory could be a very useful guide to
difficult question in the usual string theory. It requires deeper
understanding of $p$-adic string theory itself, especially of
closed $p$-adic strings (strings on $p$-adic valued worldsheet and
target space as well). It is a worthwhile task to explore $p$-adic
strings in nontrivial backgrounds. It will naturally lead to {\it
noncommutative} formulation on $p$-adic quantum theory and
examination of the corresponding Moyal-product, introduced in a
context of the noncommutative adelic quantum mechanics
\cite{kiev}. Recently, the Moyal-product has been applied in the
calculation of the noncommutative solitons in $p$-adic string
theory \cite{ghoshal}.

\section*{Acknowledgments}The research of both authors was partially
supported by the Serbian Ministry of Science and Technology
Project No 1643. A part of this work was completed during a stay
of G. Djordjevic at LMU-Munich supported by DFG Project
"Noncommutative space-time structure-Cooperation with Balkan
Countries``. Warm hospitality of Prof. J. Wess is kindly
acknowledged.

\end{document}